\documentstyle[aps,prb,twocolumn,epsf,rotate]{revtex}

\newcommand{\rr}{{\bf r}}
\newcommand{\qq}{{\bf q}}
\newcommand{\GG}{{\bf G}}

\begin{document}
\twocolumn[\hsize\textwidth\columnwidth\hsize\csname
@twocolumnfalse\endcsname

\title{Disorder and interactions in quantum Hall
ferromagnets near $\bf \nu=1$}

\draft

\author{Jairo Sinova$^{1,2}$, A. H. MacDonald$^{2}$, and S. M. Girvin$^{2}$}
\address{$^{1}$Department of Physics and Astronomy,
University of Tennessee, Knoxville, Tennessee 37996}
\address{$^{2}$Department of Physics,
Indiana University, Bloomington, Indiana 47405}
\date{\today}
\maketitle

\begin{abstract}

We report on a finite-size Hartree-Fock 
study of the competition between disorder and interactions in
a two-dimensional electron gas near Landau level filling factor $\nu=1$.
The ground state at $\nu=1$ evolves with increasing disorder 
from a fully spin-polarized ferromagnet with a charge gap, 
to a partially spin-polarized ferromagnetic Anderson insulator, to 
a quasi-metallic paramagnet at the critical point between $i=0$ and 
$i=2$ quantum Hall plateaus.  Away from $\nu=1$, the ground state evolves from a ferromagnetic 
Skyrmion quasiparticle glass, to a conventional quasiparticle glass,
and finally to a conventional Anderson insulator.  We comment on signatures of 
these different regimes in low-temperature transport and NMR lineshape and peak position
data.
\end{abstract}


\vskip2pc]
\section{INTRODUCTION}

Because of the macroscopic degeneracy of single-particle states in a Landau level,
neither disorder nor electron-electron interactions in a two-dimensional electron
system (2DES) can be treated perturbatively in the quantum Hall regime.
This is the {\it raison d'\^etre} for the many interesting and surprising 
phenomena\cite{general} which have arisen in quantum Hall physics.
Theories of quantum Hall physics usually include either only interactions or 
only disorder, although both are always present. 
In particular, it is common to include only disorder in studies of the integer quantum
Hall effect (IQHE), which generally focus on the sudden jump in the Hall conductivity 
between values separated
by $e^2/h$, and it is common to include only interactions in studies of the fractional 
quantum Hall effect (FQHE), which generally focus on the ability of interactions to 
create charge gaps at partial Landau level fillings.
The competition between interactions and disorder 
has often, but not always,\cite{Green,Nederveen}
been neglected, in part because of the lack of easily manageable
analytical and numerical tools that can deal with both simultaneously.
In this paper we address an instance in which this competition is particularly  
direct and can be successfully addressed with elementary techniques.

At Landau level filling factor $\nu=1$, the ground state of a 
disorder-free 2DES is a strong ferromagnet,\cite{Sondhi} 
{\it i.e.} it is completely spin-polarized by a Zeeman field of infinitesimal strength.
In practice, of course, the field experienced by a 
2DES in the quantum Hall regime is {\em not} infinitesimal; however 
the field's Zeeman coupling to the electron spin is typically very weak compared to  
other energy scales.  (In referring to these systems as ferromagnets
we are emphasizing that they remain spin polarized in the limit of zero Zeeman splitting.
In experimental systems,\cite{Barrett} typical values of the interaction
and Zeeman energy scales are $\sim 160 $ K and $\sim 3$ K respectively. The
spin-splitting produced by this bare Zeeman coupling is usually negligible 
in paramagnetic states.)   
The quantum Hall ferromagnet has a large gap for charge excitations, and hence
has a robust quantum Hall effect.  For typical Zeeman coupling strengths, its 
elementary charged excitations are topologically charged spin textures
(Skyrmions) containing several flipped spins.\cite{general,Sondhi,Fertig}
Large Skyrmions have a Hartree energy 
cost smaller that those of conventional quasiparticles and, because the 
spins align locally, only slightly higher exchange energy. \cite{Brey}
Because Skyrmions are the lowest-energy charged excitations, 
the global electron spin polarization is expected to decrease rapidly as $|1-\nu|$
increases. This has been observed in nuclear magnetic
resonance (NMR) experiments. \cite{general,Barrett,Fertig}
On the other hand,    
the $\nu=1$ state of non-interacting disordered electrons differs qualitatively.
The ground state is a compressible paramagnet with no Knight shift and no 
gap for charged excitations.  For zero Zeeman coupling,
quasiparticle states at 
the Fermi energy are quasi-extended and cause the Hall conductivity to 
suddenly jump by $2 e^2/h$ as this filling factor is crossed;
$\nu=1$ is in the middle of a Hall `riser', instead of being at the 
middle of a Hall plateau.

The competition between disorder and interactions at $\nu=1$ 
can be addressed\cite{olderhfdis} using the Hartree-Fock 
approximation which has the virtue of being exact\cite{general} in both the non-interacting 
and the non-disordered limits.  Experimental information on this 
competition comes primarily from transport and 
NMR studies.  Early NMR studies\cite{Barrett} of 
weak-disorder quantum Hall ferromagnets, yielded relatively featureless
lineshapes and Knight shifts in good agreement with Hartree-Fock (HF) theory
estimates\cite{Fertig,Brey,Cote} of Zeeman-coupling and filling-factor
dependent Skyrmion sizes.  (Effective field theory estimates\cite{Sondhi,Rajaraman}
are not accurate in the case of typical Zeeman coupling strengths.) 
More recent experiments\cite{Barrett_private} paint a more complex picture,
in part because the measurements were performed at lower temperatures 
where the signal is not motionally averaged.\cite{Jairo,Yalegrouppapers}
It is now clear that disorder plays a role in the interpretation 
of these experiments, even when it is weak. 
At stronger disorder, as the non-interacting limit is approached,
the spin-polarization must eventually
vanish.  In our model calculations we find that 
as the interaction strength is increased relative to disorder at $\nu=1$, 
the 2DES ground state suffers a continuous phase transition a from paramagnetic to 
a ferromagnetic state.  Depending on the details of the disorder model,
a second continuous phase transition to a fully spin-polarized
incompressible strong ferromagnet 
with a gap for charged excitations may occur at still stronger interactions.
For the disorder models we use, the fully polarized state is reached 
when the Coulomb energy scale is approximately twice the Landau-level-broadening
disorder energy scale.  Away from $\nu=1$, screening by mobile charges 
reduces the importance of disorder and the system reaches
maximal spin-polarization at smaller interaction strengths.
The maximally polarized ground state at moderate interaction strengths is 
best described as a glass of localized conventional quasiparticles
formed in the $\nu=1$ fully polarized vacuum.  Only for stronger interactions 
do we find a phase transition to a state with non-collinear magnetization in which
the localized particles have Skyrmionic character.

We organize this paper as follows. In Sec. II we summarize our implementation
of finite-size HF theory in the lowest Landau level (LLL).  In Sec. III and IV we present 
and discuss our numerical results for calculations at $\nu=1$ and $\nu\ne 1$.
The possibility, discussed in recent work,\cite{Nederveen} that at $\nu=1$ disorder will induce
reduced-size Skyrmion-anti-Skyrmion pairs in the ground state is specifically
addressed in Sec. III.
Finally, in Sec. V we present our conclusions.

\section{Hartree-Fock Theory in the LLL}

The HF approximation allows the interplay between disorder and
interactions to be addressed while 
retaining a simple independent-particle picture of the many-body ground state.
For the current study, the use of this approximation is underpinned by 
the fact that it reproduces the exact ground state at $\nu=1$ in 
both weak interaction and strong interaction\cite{general} limits.
HF theory is a self-consistent mean field theory, 
and as such it has strengths and shortcomings, which we discuss later.  
In this section, we outline the basic formalism of 
HF approximation calculations in the LLL limit.

In a strong magnetic field, the Landau level splitting is very large and,
since excitations to higher Landau levels are effectively forbidden at the 
low experimental temperatures, we follow the common practice of 
considering only LLL states. 
Neglecting the frozen kinetic-energy degree of freedom, the Hamiltonian
in second quantization is written as
\begin{equation}
{\cal H}={\cal H}_I+{\cal H}_{dis}+{\cal H}_Z\,,
\end{equation}
where ${\cal H}_I$ is the interaction part of the Hamiltonian
\begin{eqnarray*}
{\cal H}_I&=&\frac{1}{2}\int d\rr \int d\rr' \sum_{\sigma\, \sigma'}
v_I(\rr-\rr') \hat\psi^\dagger_\sigma(\rr)
\hat\psi^\dagger_{\sigma'}(\rr') \hat\psi_{\sigma'}(\rr') \hat\psi_\sigma(\rr)\,\,,
\end{eqnarray*}
${\cal H}_{dis}$ is the external disorder part of the Hamiltonian
\begin{eqnarray*}
{\cal H}_{dis}&=&\int d\rr\sum_{\sigma}v_{E}(\rr)
\hat\psi^\dagger_\sigma(\rr)\hat\psi_\sigma(\rr)\,,
\end{eqnarray*}
and ${\cal H}_Z$ is the Zeeman term
\[
{\cal H}_Z=-\frac{1}{2}g\mu_B \int d\rr \sum_{\sigma \sigma'} 
\hat\psi^\dagger_{\sigma'}(\rr') \hat\psi_{\sigma}(\rr')
\vec{\tau}_{\sigma' \sigma}\cdot\vec{B}(\vec r)\,,
\]
with $\sigma=\uparrow,\downarrow$, $v_I$ and $v_E$ being the Coulomb interaction
and disorder potentials respectively, and $\tau_i$ being the Pauli matrices. 
Here we also define  the Zeeman coupling strength
as $\tilde g=g\mu_B B/(e^2/\epsilon l)$ for later reference.
We chose the the Landau gauge  elliptic theta functions as our basis
\[
\phi_m(x,y)=\frac{1}{\sqrt{L_y l \sqrt{\pi}}}
\sum_{s=-\infty}^\infty e^{i\frac{1}{l^2}x_{m,s}y}
e^{-\frac{1}{2 l^2}(x-x_{m,s})^2}\,\,,
\]
where $x_{m,s}=\frac{2\pi m l^2}{L_y}+s L_x$,
$m,m'=1,\dots,N_\phi$, $N_\phi=A/(2\pi l^2)$, and $l$ is
the magnetic length which we set equal to $1$ for simplicity.  
These wavefunctions satisfy
the semi-periodic boundary conditions $\phi_m(x,y)=\phi_m(x,y+L_y)$ and
$\phi_m(x+L_x)=\exp(+iL_x y/l^2)\phi_m(x,y)$.

We consider the HF single particle states to be a linear combination
of the up and down spin states of these basis functions
\[|\alpha\rangle=\sum_{m,\sigma}\langle m \sigma|\alpha\rangle \,\,
|m \sigma\rangle\,\,.\]
Before writing down the Hamiltonian matrix in the HF approximation, we 
introduce several notation simplifying definitions, closely following 
previous HF studies. \cite{olderhfdis}
The expectation value of the particle density (in momentum space) is given by
\begin{eqnarray*}
\langle \rho(\qq)\rangle&=&\sum_\alpha n_F(\epsilon_F-\epsilon_\alpha)\langle \alpha
|e^{-i\qq\cdot\rr}|\alpha\rangle\\
&\equiv&N_\phi e^{-\frac{1}{4}q^2}
\sum_{\sigma,\sigma'} \delta_{\sigma,\sigma'} \Delta_{\sigma'\,\sigma}(\qq)
e^{-i\frac{q_x q_y}{2}}\,\,,
\end{eqnarray*}
where we define
\[
\Delta_{\sigma'\,\sigma}(\qq)\equiv\frac{1}{N_\phi}
\sum_{m,m'}\delta_{(x_{m'},x_m+q_y)}e^{-iq_x x_m}
\rho_{\sigma'\,\sigma}(x_{m'}|x_m),
\]
and
\begin{eqnarray*}
\rho_{\sigma'\,\sigma}(x_{m'}|x_m)
&=&\sum_\alpha n_F(\epsilon_F-\epsilon_\alpha) \langle
m'\sigma'|\alpha\rangle\langle\alpha|m\sigma\rangle\,.
\end{eqnarray*}
Here $\delta_{(x_{m'},x_m+q_y)}$ is a periodic Kronig delta function,
{\it i.e.} it is nonzero for $x_{m'}=x_m+q_y+s\,L_x$ for any integer s.
With these definitions we can  write the Hamiltonian matrix in the
HF approximation in a compact form which is simple to diagonalize numerically
\begin{eqnarray}
&&\langle m \sigma|{\cal H}|m' \sigma'\rangle_{HF} =
\sum_{\qq\epsilon {\rm BZ}}
\left\{(\gamma\Delta_0(\qq){\rm U_H}(\qq)+{\rm U_{\rm dis}}(\qq)) {\rm I}+\right.
\nonumber\\
&&\left.\frac{\gamma}{2}(\Delta_0(\qq){\rm I}+\vec{\Delta}(\qq)\cdot\vec{\tau})
{\rm U_F}(\qq)\right\}
\delta_{(x_m,x_{m'}+q_y)}e^{+iq_x x_{m'}}\nonumber\\
&&-\frac{1}{2}\tilde g \hat B\cdot \vec{\tau}\,,
\end{eqnarray}
where $\Delta_\alpha(\qq)={\rm Tr}\{\Delta_{\sigma \sigma'}(\qq)
\tau^{\alpha}\}$, and $\hat B$ specifies the orientation of the
external magnetic field. 
The various effective potentials which appear here are defined as
\begin{equation}
{\rm U_{dis}}(\qq)=\frac{1}{A}\sum_\GG
e^{-\frac{1}{4}|\qq+\GG|^2}v_E(\qq+\GG) e^{
\frac{i}{2}(q_x+G_x)(q_y+G_y)}\,\,,
\label{Udis}
\end{equation}
\begin{equation}
{\rm U_H}(\qq)=\frac{1}{2\pi}\sum_{\GG} e^{-\frac{1}{2}|\qq+\GG|^2}
\frac{2\pi e^2}{|\qq+\GG|}(1-\delta_{\qq+\GG,0})\,\,,
\end{equation}
and
\begin{equation}
{\rm U_F}(\qq)=-\frac{1}{A }\sum_{\qq'}
e^{-\frac{1}{2}|\qq'|^2}e^{i{q'}_x q_y-iq_x q'_y}
\frac{2\pi e^2}{|\qq'|}(1-\delta_{\qq',0})\,,
\label{U_F}
\end{equation}
with $\GG=(\frac{2\pi N_\phi}{L_x} n_x,\frac{2\pi N_\phi}{L_y} n_y)$.

For $v_E(\rr)$, we choose a white noise potential without spatial correlation,
$\langle\langle v_E(\rr)v_E(\rr')\rangle\rangle=\sigma^2 \delta(\rr-\rr')$.
The density of states in the non-interacting limit has been calculated
exactly by Wegner for this distribution of the
disorder potential,\cite{Wegner} yielding a full width at half maximum of
approximately $1.06 \sigma$.
In our calculations the parameter $\gamma=e^2/\epsilon\sigma$  specifies 
the relative strength of interactions and disorder broadening. 
This type of disorder potential distribution is characterized by a single 
parameter $\sigma$, which we use as our unit of energy.
As we discuss later, our results are insensitive to correlations in
the disorder potential on length scales smaller than $l$, but would change
in some respects for disorder potentials which are smooth on the 
magnetic length scale. 

The HF equations are solved by an iterative approach which
can create difficulties which must be addressed.  The HF equations
generally have many solutions that correspond to different, usually
metastable, extrema of the HF energy functional. The challenge
is to locate the true global minimum.  In particular, the iteration
process will not break any symmetries of the Hamiltonian which are 
not broken by the starting charge and spin-densities, even though
the global minimum of the HF energy functional frequently 
does break at least some of these symmetries.
To counter such 
problems, it is usually a good idea to introduce small artificial terms in
the Hamiltonian which break the continuous symmetries and help the iterative
process to reach the lowest energy state. 
In this problem, the iterative process
is also hampered by severe convergence problems at zero temperature connected
with the localization of HF quasiparticle wavefunctions and 
the long range nature of the Coulomb interactions.  A small change
in the energy of a particular orbital may involve a substantial rearrangement of the
charges.  These problems can be mitigated by always working at a 
temperature which is comparable to the finite-size quasiparticle energy level spacing
and which scales to zero as the system size increases.
The Zeeman term in the Hamiltonian, for which we choose typical experimental values,
reduces the SU(2) spin symmetry to a U(1) symmetry.  In order to 
break the continuous U(1) symmetry we introduce an artificial (but very
small) local magnetic field at the center of our simulation cells which 
points in the x-direction.  It is this space-dependent field, required on 
purely technical grounds, which has motivated developing
the formalism in a manner which permits non-constant Zeeman coupling
strengths.  To ensure that this artificially field and the finite 
temperature do not affect the final solution, we lower the
magnitude of these terms until no change is seen in 
local charge and spin densities, or in HF quasiparticle energies.

The phase diagrams discussed in Secs. III and IV, were obtained by 
starting from the non-interacting case and incrementing the interaction strength $\gamma$,
taking as the starting densities the self-consistent densities from the previous 
$\gamma$ value. There is, of course, some hysteresis involved in this 
process so we do a backwards sweep on $\gamma$ once we have reached the maximum
interaction strength for a given run.  If they differ, we use the smaller of the 
values obtained in upward and downward sweeps for the energy per particle.
The energy per particle is obtained using the expression
\begin{eqnarray*}
&&\frac{E}{N}=\frac{1}{\nu}\sum_{\qq\in {\rm BZ}}{\rm U}_{\rm dis}(\qq) {\Delta_0}^*(\qq)
+\frac{\gamma}{2\nu}\sum_{\qq\in {\rm BZ}}{\rm U}_{\rm H}(\qq) |\Delta _0(\qq)|^2\\
&&+{\rm U}_{\rm F}(\qq)
\left(|\Delta _{\uparrow\uparrow}(\qq)|^2
+|\Delta _{\downarrow\downarrow}(\qq)|^2+|\Delta_{\uparrow
\downarrow}(\qq)|^2+| \Delta_{\downarrow\uparrow}(\qq)|^2\right)\\
&&-\frac{\gamma}{2} \tilde g \hat B\cdot \vec{P}_{tot}\,,
\end{eqnarray*}
where $ \vec{P}_{tot}$  is the total global spin polarization.
The local spin magnetization density, which we calculate as well, is given by 
\[
\langle S_i(\rr)\rangle=
\frac{\hbar}{2}
\sum_{m',m,\sigma,\sigma'} \rho_{\sigma \sigma'}(x_{m'}|x_m)\tau^i_{\sigma \sigma'}
\phi_m^*(\rr)\phi_{m'}(\rr)\,\,.
\]
We define the local spin polarization as 
$\langle P_i(\rr)\rangle=2\langle S_i(\rr)\rangle/(\hbar 
\langle \rho(\rr)\rangle)$.
Note that in this case $\langle |\vec P(\rr)| \rangle$ does not have
to be equal to $1$ since the system is compressible except
in the limit of very large $\gamma$ where
$\langle |\vec{P}(\rr)|\rangle\rightarrow 1$ for all $\rr$.

Our criteria for convergence is that 
\begin{equation}
\delta \Delta\equiv\frac{1}{N_\phi^2} \sum_{\qq \in {\bf BZ}}
\sum_{\sigma \sigma'} 
|\Delta_{\sigma \sigma'}^{i}(\qq)-\Delta_{\sigma \sigma'}^{i-1}(\qq)|^2< 1\times
10^{-6}\,,
\end{equation}
where $i$ stands for the $i$th iteration.
We have performed calculations for several disorder realizations
at different values of $\nu$ for system sizes of $N_\phi=16-32$.
The finite size effects come mainly from the effective exchange
potential $U_F(\qq)$ and have been studied in detail previously.\cite{Allan_finite}
The main effect is on the interaction part of the energy per particle
and it is well understood and easily corrected.  We believe that the physics of
the phase transitions observed in these calculations is not 
affected qualitatively by finite-size effects.
Our qualitative conclusions are based on 
persistent features which are obtained for several different disorder realizations.
The values of $\gamma$ at which
the various transitions and cross overs we discuss below take place, do not
change by more than $5\%$ for different realizations. 
The results we present here are for one particular disorder realization. 

\section{Results at $\nu=1$}
\begin{figure}
\epsfxsize=3.375in
\centerline{\epsffile{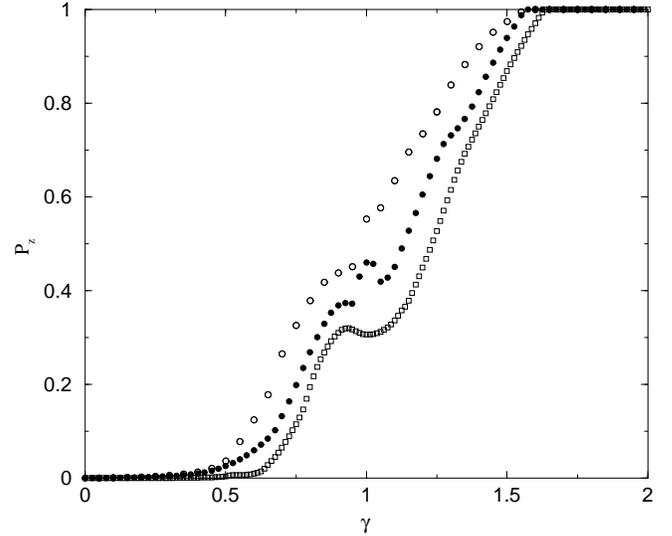}}
\caption{Global polarization for a fixed disorder realization
at $\nu=1$
with  $\tilde g =0.015$ (filled circles) and $\tilde g =0.0018$ (squares).
The average {\em magnitude} of the local polarization is shown (open circles) at
$\tilde{g}=0.015$.
The transition to a fully polarized state occurs at $\gamma\approx \,1.6-1.9$
for all disorder realizations studied. 
Examples of the local polarization density and 
local density profile at different values of $\gamma$ for a typical  
disorder realization are shown in Figs. \ref{local_pol2}
and \ref{local_dens2}.}
\label{phase_diag2}
\end{figure}

At $\nu=1$ the disorder free ($\gamma \to \infty$) 2DES has\cite{general} a 
$S =N/2$ ground state.  The $S_z=S=N/2$ member of this multiplet is a single
Slater determinant and can therefore be obtained by solving Hartree-Fock
equations self-consistently.
It is only in recent years that samples which are sufficiently clean to 
reach, or at least nearly reach, complete spin polarization have been grown.\cite{Barrett} 
The collective behavior producing such a ground state was not exhibited in
earlier samples which had more disorder in the form of 
unintended impurities, interface dislocations, and,
in modulation doped samples, the potential from remote ionized 
donors.  Fig. \ref{phase_diag2} summarizes the HF theory results we
have obtained for the dependence of the spin polarization on interaction strength.
The calculations were performed for a realistic value of the Zeeman coupling
strength, $\tilde g=0.015$, and at a very small value, $\tilde g=0.0018$.  
Extrapolating from these two values to $\tilde g =0$, allows us to 
identify parameter values for which spontaneous spin polarization occurs,
{\it i.e.,} values for which the ground state is ferromagnetic.
We find that ferromagnetism occurs for $\gamma \gtrsim 0.5$ in the 
Hartree-Fock approximation; at smaller values of $\gamma$ the single-particle
disorder term dominates and yields a spin-singlet ground state.  Notice that 
the spin susceptibility, which may be estimated from the difference between
the spin-polarizations at the two $\tilde g$ values, is small in the singlet
state, and becomes large as the phase transition to the ferromagnetic state is
approached.  For the specific finite-size disorder realization we have
studied, complete spin polarization is reached at a finite 
value of $\gamma \sim 1.5$.   At larger values of $\gamma$, the system 
has a finite gap for charge excitations.  We must be aware,
however, that the HF approximation overestimates the tendency 
of the system to order so
the interaction strength at both transition points should be taken as lower limits.
In addition, any physically realistic disorder potential is likely to have 
rare strong disorder regions which prevent the fully polarized state from 
being reached.

\begin{figure}
\epsfysize=6.75in
\epsfxsize=3.375in
\centerline{\epsffile{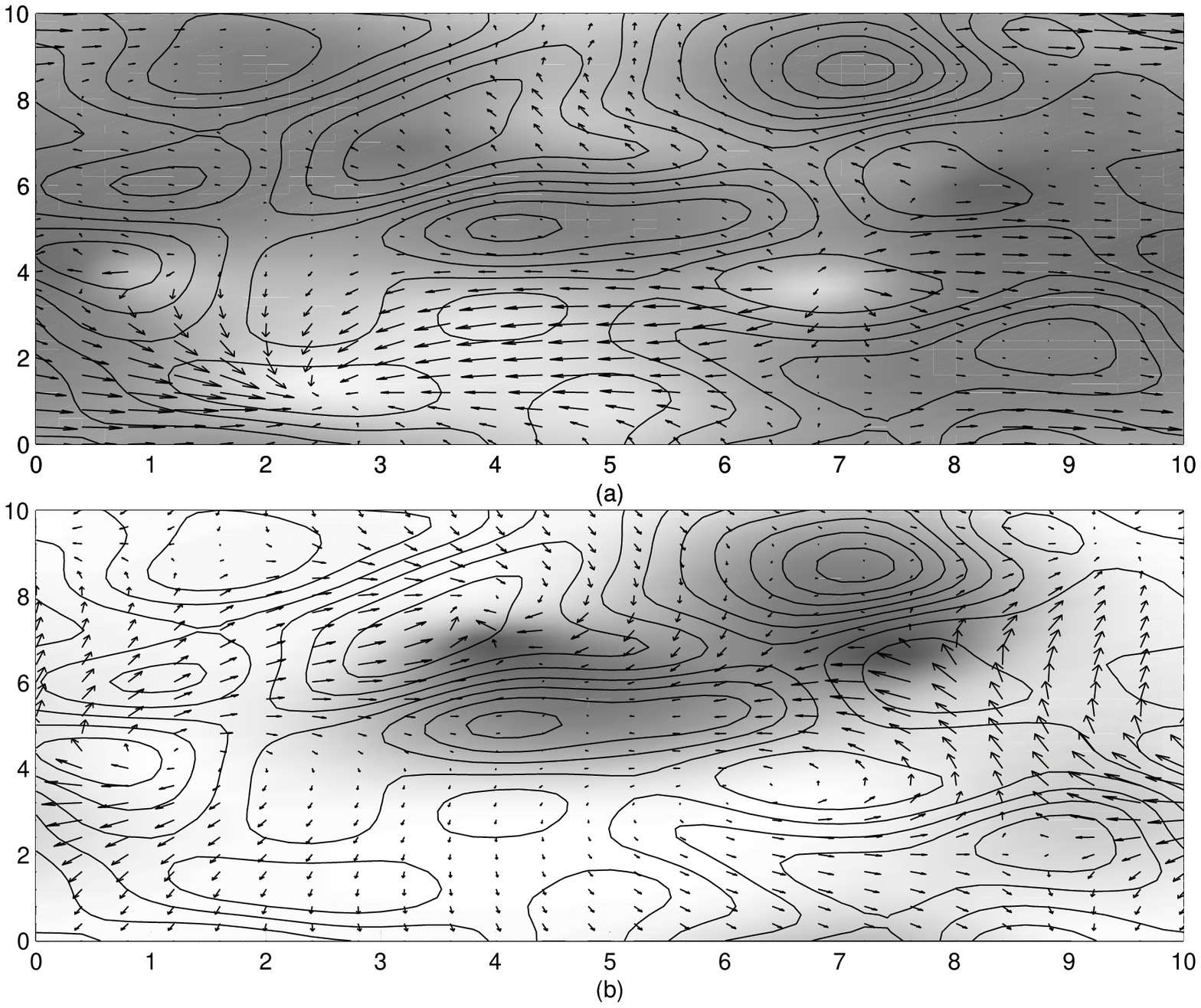}}
\caption{Local polarization density for $\nu=1$  and
$\tilde g =0.015$ at $\gamma=0.75$ (a) and
at $\gamma=1.25$ (b). The z-component is represented by a shadow plot with
black as $-1$ and white as $+1$. The in-plane local polarization density
is represented by a two-dimensional
vector plot. The contour plot corresponds to the effective disorder
potential used in the phase diagram shown in Fig. \ref{phase_diag2}.}
\label{local_pol2} 
\end{figure}

\begin{figure}
\epsfysize=6.75in
\epsfxsize=3.375in
\centerline{\epsffile{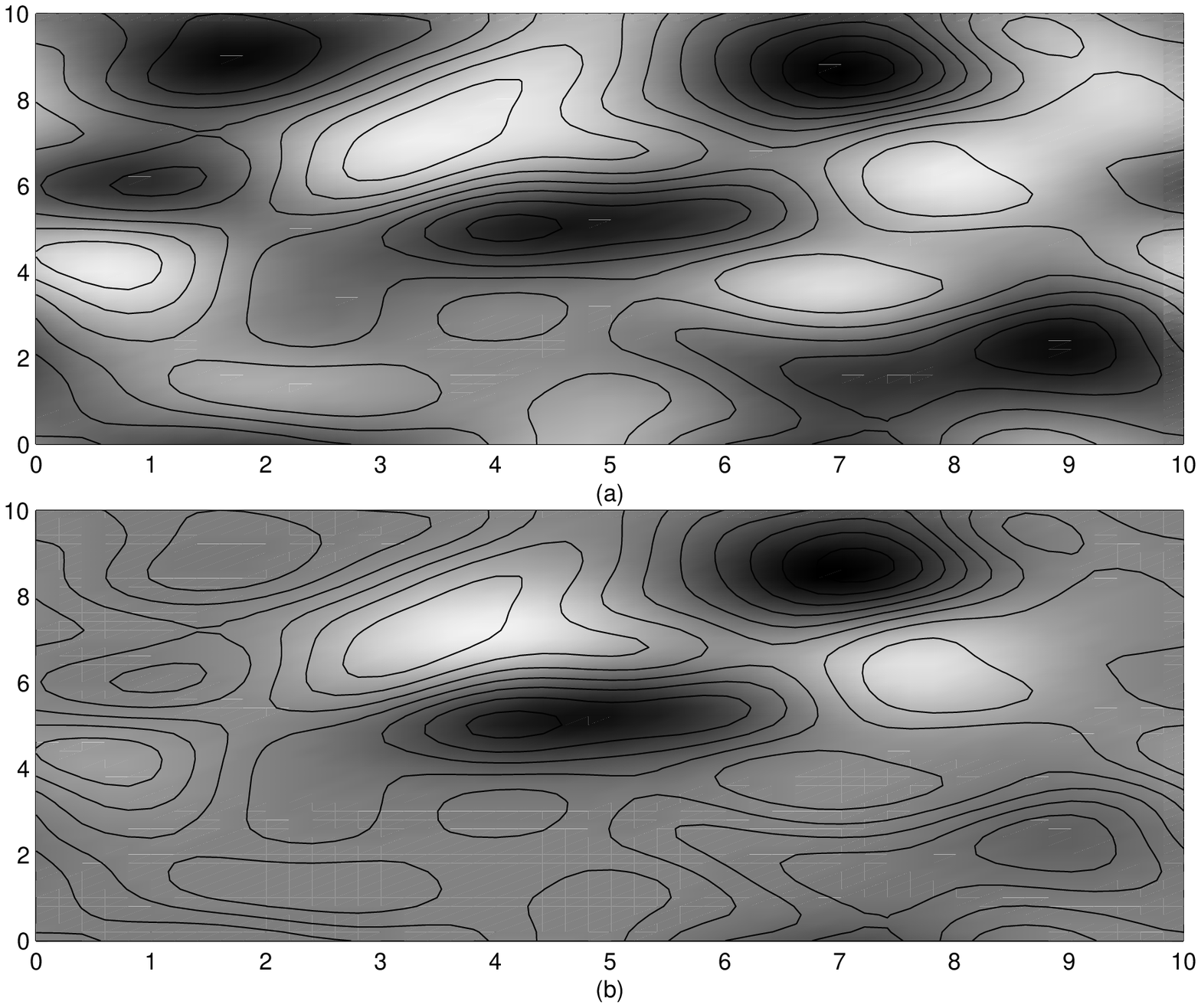}}
\caption{Local density profile for $\nu=1$  at $\gamma=0.75$ (a) and
at $\gamma=1.25$ (b).  In this shadow plot, black
represents a local Landau level filling factor of two and 
white represents a local Landau level filling factor of zero.
The contour plot corresponds to the effective disorder
potential used also in the calculations of Fig. \ref{phase_diag2}.
Minima in the electron density occur at maxima in the effective potential
and vice versa.
}
\label{local_dens2} 
\end{figure}

In our calculations, there is a wide region of 
interaction strengths $\gamma$ for which partially spin-polarized states occur.
In this regime our HF ground states nearly always have non-collinear magnetic order.
We show local spin polarization and charge density profiles 
of typical partially polarized states in Figs. \ref{local_pol2} and \ref{local_dens2}.
The origin of the reduced spatially integrated spin polarization is partly due to 
variation of spin-orientation, but principally due to a reduction in the average 
value of the of the {\em magnitude} of the local spin polarization. 
This point is illustrated in Fig. 1 (open circles) and may be inferred from
Fig.  \ref{local_pol2} (a).  The reduction in spin-polarization is due to the 
occurrence of doubly-occupied orbitals, {\it i.e.} to disorder induced charge 
fluctuations which cannot be accurately described in models which include
only the spin degree of freedom.  
The charged excitations of the ground state in this regime are 
ungapped and involve population of localized quasiparticle states.
We also remark that local density profiles at these relatively
small $\gamma$ values, illustrated in in Fig. \ref{local_dens2} (a), follow
the effective disorder potential smoothed by the form factor for 
lowest Landau level electrons.  Rapid spatial variation components in  
the white noise model disorder potential have little effect on the 
electronic state.  The relationship between electron number density and 
the Pontryagan index density of the local spin orientation, valid 
for slow spin-orientation variation and nearly constant charge density,\cite{general}
is {\em not} valid in this regime.  Still, 
the collective nature of the 2DES manifests itself in the nonzero  
spin polarization density perpendicular to the Zeeman field. 
As interactions strengthen further, the local charge density smoothes out
favoring the minimization of Coulomb energy at a cost in 
disorder energy.  (See  Fig. 2 (b) and 3 (b) for $\gamma\approx 1.5$.)

\begin{figure}
\epsfxsize=3.375in
\centerline{\epsffile{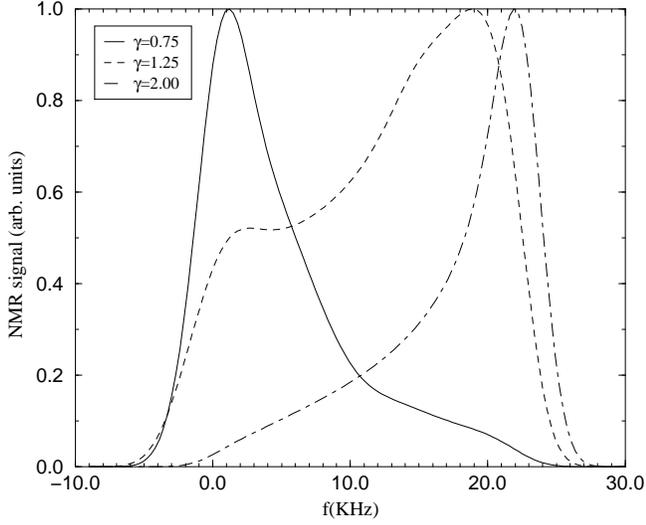}}
\caption{NMR spectrum for $\nu=1$ and $\tilde g =0.015$ at $\gamma=0.75$ (solid line), at
$\gamma=1.25$ (dashed thick line), and at $\gamma=2.0$ (dashed-dotted line).
The sample parameters correspond to ones the used in the experimental
studies of Ref. [13].}
\label{NMR2} 
\end{figure}

Experimentally, the effects of disorder can be seen most directly in the NMR
spectral line shape obtained at the lowest possible temperatures where the spin profile
is frozen on the experimental time scale. 
\cite{Jairo,Yalegrouppapers} The NMR intensity spectrum in this regime is
given by
\begin{equation}
I(f,\gamma)\propto \int d{\bf r} \rho_N(z)e^{
-\frac{1}{2\sigma^2}(2\pi f -2\pi K_s \rho_e(z) \langle \vec{S}(\rr;\gamma)\rangle)}
\,,
\label{NMRspectra}
\end{equation}
with $\sigma=9.34 {\rm ms}^{-1}$ and $K_s\sim 25{\rm KHz}$.
Here $\rho_N(z)$ is the nuclear polarization density 
and $\rho_e(z)$ is the electron density envelope function in the quantum well.
The evaluation of such spectra has been outlined elsewhere;\cite{Jairo,Yalegrouppapers}
here we simply show results for several interaction 
strengths in Fig. \ref{NMR2}. 
The parameters used in Eq. \ref{NMRspectra} are the same as the
ones used in Ref.\ \onlinecite{Jairo}. 

Note that the quantity usually identified experimentally as the 
Knight shift, the location of the peak in the NMR spectrum
in Fig. \ref{NMR2}, does not match the global polarization.  This Knight shift measurement 
always overestimates the global polarization.
In order to obtain the global 
polarization of the system from the measured spectrum one has to extract 
the first moment of a normalized spectrum.\cite{Jairo}
One sees from the NMR spectrum at $\gamma=1.25$ that disorder induced 
spin density variation leads to a broadening  of the maximum peak
and can even lead to secondary peaks at lower Knight shift frequencies.
Note, however, that features corresponding to negative Knight shifts,
corresponding to regions of reversed electronic spins, are 
unlikely because the typical size of such
regions is small and because they are also obscured by the finite width
of the quantum wells which trap the 2DES.
Our calculations demonstrate that care must be taken in interpreting 
low temperature NMR data in the quantum Hall regime.

The partially polarized regime can also be studied experimentally by 
measuring the transport activation gap.  Provided that weak 
Zeeman coupling can be ignored, the extended quasiparticle states
are expected to be precisely at the Fermi level when the 2DES 
is in a paramagnetic state.  The Hall conductivity should jump from
$0$ to $2 e^2/h$ at $\nu=1$. In the ferromagnetic state,
the majority-spin extended quasiparticle state will be below 
the Fermi level, the majority-spin extended state will
be above the Fermi level and the Hall conductivity at $\nu=1$
should be quantized at $\sigma_{xy} = e^2/h$.  The spontaneous
splitting of the two extended state energies is experimentally
accessible and should exhibit interesting power law critical 
behavior as the ferromagnetic state is entered.   This transport gap 
should vary monotonically with the global spin polarization, although
the precise relationship between these quantities is not trivial.

It is interesting to note that the Skyrmion-anti-Skyrmion pairs predicted recently 
by Nederveen and Narazov \cite{Nederveen}
do not appear in our calculations.  We do not conclude that 
these objects cannot appear at $\nu=1$; we would expect them, for
example, if we choose a disorder model with relatively large potential variations,
but only on a length scale much larger than the Skyrmion size.  
In this case the NL$\sigma$ model considerations in Ref.\onlinecite{Nederveen}
should be applicable.  Our calculations demonstrate rather clearly
however, that charge density variation at $\nu=1$ does not 
necessarily, or even usually, require the existence of well defined Skyrmion
quasiparticles.

\section{Results at $\nu\ne 1$}

In clean (large $\gamma$) samples where full polarization is observed at $\nu=1$, 
the global polarization decays rapidly with $|1-\nu|$.\cite{Barrett}
It is generally accepted that this property is a unique signature 
which experimentally establishes the thermodynamic stability of 
Skyrmion collective quasiparticles. 
In the strong disorder limit, on the other hand, 
spontaneous spin polarization does not occur 
at any filling factor near $\nu = 1$.

\begin{figure}
\epsfxsize=3.375in
\centerline{\epsffile{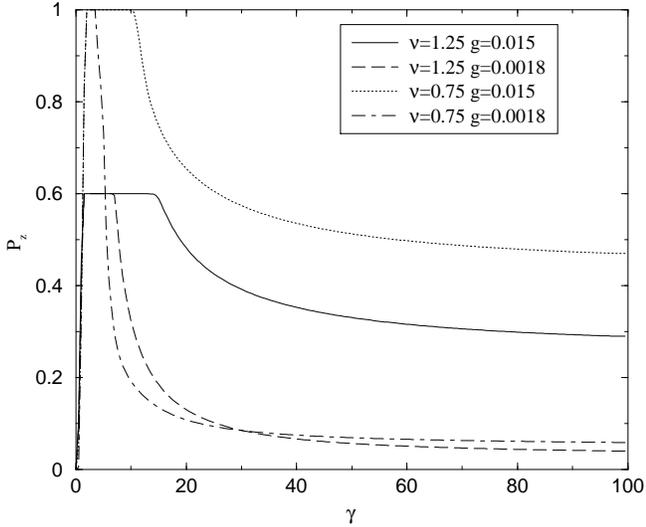}}
\caption{Global polarization phase diagram for a fixed disorder realization
at $\nu=1.25$ and $\tilde g =0.015$ (solid line),
at $\nu=1.25$ and $\tilde g =0.0018$ (dashed line),
at $\nu=0.75$ and $\tilde g =0.015$ (dotted line),
and at $\nu=0.75$ and $\tilde g =0.0018$ (dotted-dashed line).
The transition to a maximally polarized state 
(Laughlin quasiparticle glass) occurs at $\gamma\approx \,1.5-2$
for all disorder realizations obtained. The transition from a 
Laughlin quasiparticle glass to a Skyrmion glass occurs at $\gamma\approx 15$
for experimentally relevant parameters.
Local spin polarization density and 
local density at different values of $\gamma$ for such
disorder realization are shown in Figs. \ref{local_pol3}
and \ref{local_dens3}.}
\label{phase_diag3}
\end{figure}

The global polarization results for $\nu \ne 1$ 
in Fig. \ref{phase_diag3}, illustrate how the system interpolates
between these two extrema. As the interaction strength $\gamma$ is increased from 0 to
2, the behavior is similar to the $\nu=1$ case.  For strong disorder
charge variation is dominant, and small spin polarizations occur 
primarily because many single particle orbitals are occupied by 
both up and down spin electrons.
Charge variation is the dominant response to disorder, 
and it continues to play an important role at all interaction strengths.
At sufficiently large $\gamma$,   
our finite size systems reach a state with the maximum spin 
polarization allowed by the Pauli exclusion principle.
This  maximally polarized state is reached earlier than in the case
at $\nu=1$ ($\gamma\sim1.4-1.6$) because, we believe, a larger number of 
charged quasiparticles are available to screen the random potential.
At this point the system forms what we refer to as a conventional
quasiparticle glass (CQG).  The 
conventional Laughlin quasiparticles are initially localized in the deepest 
minima (or maxima for $\nu<1$)  of the 
effective disorder potential and as the interaction strength increases,
or equivalently the depth of the disorder potential wells becomes 
smaller, the charged quasiparticles rearrange themselves locally 
into a quasi-triangular Wigner crystal pinned by the strongest of the disorder potential
extrema. At larger $\gamma$ we observe a transition from a CQG to a Skyrmion glass.
The location of this transition is marked by a reduction of the
global polarization from its maximally polarized value.
For a specific disorder realization the point of cross over from the CQG to
the Skyrmion glass, as illustrated in Fig. \ref{phase_diag3},
depends on filling factor and $\tilde g$.
The dependence of the transition point on $\tilde{g}$ in this regime
can be approximated by considering a simple model for a single Skyrmion trapped
at a disorder potential extrema.  We approximate its energy by 
\begin{equation}
E(K)=U(K-K_0)^2+g^*\mu_B B K+\sigma AK\,,
\label{model}
\end{equation}
where  $K$ is the number of spin flips per Skyrmion, $\sigma$ is the 
strength of the disorder potential and $A$ is a phenomenological 
parameter.  The first two terms
determine the optimal Skyrmion size in the absence of disorder.\cite{UKmodel_Allan}
The form for the third term reflects the property that Skyrmions with smaller 
$K$ are smaller and will be able to concentrate more strongly close to the 
potential extrema.  This simple model 
gives an estimate of the interaction strength at which $K > 0$ Skyrmions
first become stable 
\begin{equation}
\gamma^*=\frac{A}{2K_0U/(e^2/\epsilon \ell)-\tilde{g}}\,.
\label{gammastar}
\end{equation}
The parameters $U$ and $K_0$ can be estimated\cite{UKmodel_Allan} for filling factor $\nu=1.25$
as $U/(e^2/\epsilon \ell)\sim0.014$ and $K_0\sim 1$. Using
our numerical result for where the transition occurs at $\tilde{g}=0.0018$, we 
estimate that $A\sim 0.1$. From this, one obtains an 
estimate of $\gamma^*\sim 7$ for the cross over
point from conventional quasiparticles to Skyrmions at $\tilde{g}=0.015$. 
This is in reasonable agreement with the actual cross over point
$\gamma^*\sim 10$ (see Fig. \ref{phase_diag3}, the transition is out of 
scale in Fig. \ref{phasediag}) given the simplicity of the model.  These estimates of the 
maximum disorder strength at which Skyrmion physics is realized could
be checked by performing NMR experiments in samples where 
electron density, and hence the interaction strength, is adjusted 
by the application of gate voltages.

\begin{figure}
\epsfysize=6.75in
\epsfxsize=3.375in
\centerline{\epsffile{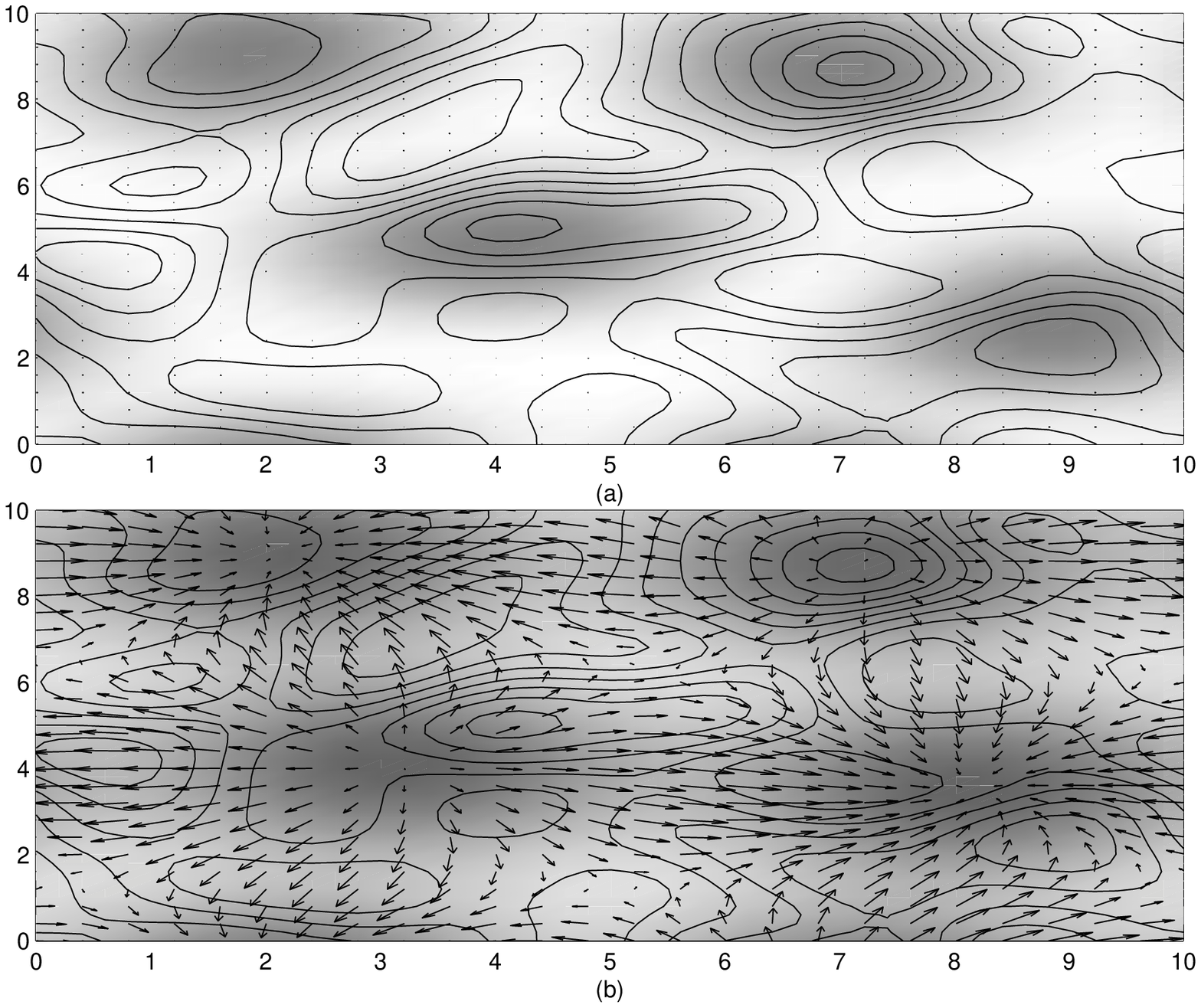}}
\caption{Local spin-polarization density for $\nu=1.25$ at $\gamma=10.0$ (a) and
at $\gamma=40.0$ (b). The z-component of the spin is represented by a shadow plot with
black as $-1$ and white as $+1$. The in-plane local spin-polarization density
is represented by a two-dimensional vector
plot. The contour plot shows the specific effective disorder
potential which leads to these results and those illustrated in 
Fig. \ref{phase_diag3}.  Notice that spontaneous in-plane spin-polarization 
appears only at the larger $\gamma$ value.}
\label{local_pol3} 
\end{figure}

\begin{figure}
\epsfysize=6.75in
\epsfxsize=3.375in
\centerline{\epsffile{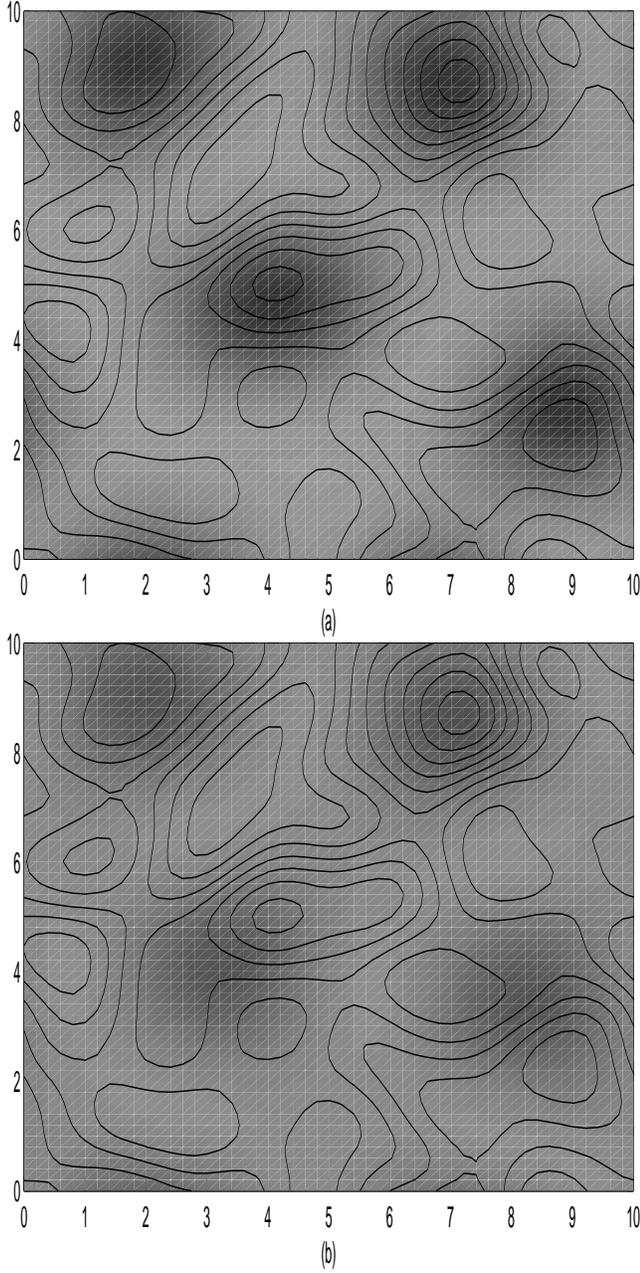}}
\caption{Local density for $\nu=1.25$  at $\gamma=10.0$ (a) and
at $\gamma=40.0$ (b).  The density is represented by a shadow 
plot where black correspond to local filling factor $\nu=2$ and 
white corresponds to local filling factor $\nu=0$.  
The contour plot shows the effective disorder
potential used in this calculation and in Fig. \ref{phase_diag3}.
Notice that the density variation is smoother at larger $\gamma$ 
when Skyrmion quasiparticles, rather than Laughlin quasiparticles,  are 
localized near potential minima. }
\label{local_dens3} 
\end{figure}

For a particular realization of the disorder potential, particle-hole
symmetry is broken in a finite system and is recovered only  
in the limit of very large $\gamma$.  The 
particle-hole symmetry relation for the global spin polarization is 
i.e. $(1-\epsilon) P_z(\nu=1-\epsilon)=(1+\epsilon) P_z(\nu=1+\epsilon)$ where $\epsilon
<1$.  At large $\gamma$ this relation is approximately satisfied.
Also in this limit, the Pontryagan relation between the local density
profile and the local spin density \cite{general} becomes accurate.
In the clean limit, the Skyrmion system crystallizes in a square 
lattice for the filling factors considered here.  (The Skyrmion crystal 
is triangular\cite{UKmodel_Allan} for $\nu$ very close to 1.)  The disordered Skyrmion glass
state has very smooth fluctuations of the local spin density, compared to the
CQG, although both lattices are pinned by the disorder
potential.   We remark that quantum fluctuations in Skyrmion positions are not
accounted for in HF theory, and it  is quite possible that even in this limit the 
ground state is a liquid rather than a crystal. \cite{Juanjo}
As noted in Ref. \onlinecite{UKmodel_Allan} it is possible that the broken U(1)
symmetry of the Skyrmions orientation order predicted by Hartree-Fock does not
survive quantum fluctuations.
We show an example of the CQG in Fig. \ref{local_pol3} (a) and
\ref{local_dens3} (a), and of the quasi Skyrmion lattice state in Fig. \ref{local_pol3}
(b) and \ref{local_dens3} (b).   Note that we find, in agreement with Nederveen
and Narazov, \cite{Nederveen} a shrinking of the Skyrmion size as
disorder broadening increases.
This effect may help explain the appearance of a ``tilted plateau'' centered around $\nu=1$ 
in the Knight shift vs. filling factor data. \cite{Barrett_private} 
Rare highly  disorder regions in the sample may localize and
reduce the effective size of  the few Skyrmions present at 
these filling factors. This would allow the  bulk of the sample  to be fully polarized 
at $\nu\ne 1$ and give rise to a Knight shift equivalent to the one at $\nu=1$. 
The plateau is tilted because of the change in fully polarized density
as pointed out in Ref. \onlinecite{Barrett_private}.

\section{Discussion}

\begin{figure}
\epsfxsize=3.375in
\centerline{\epsffile{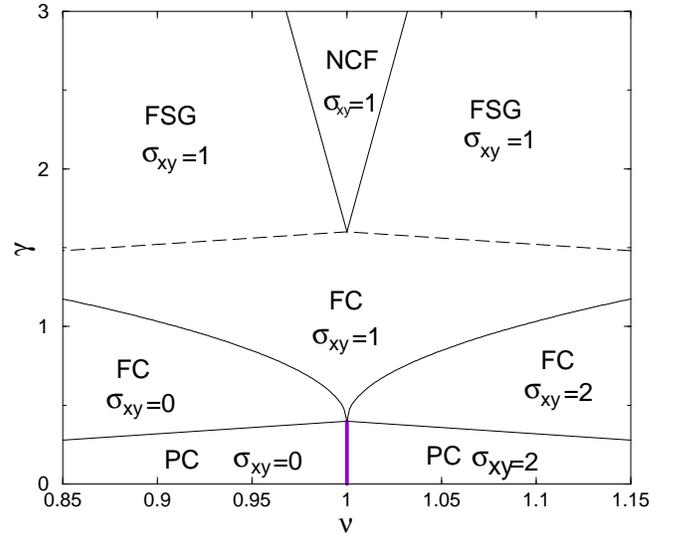}}
\caption{Phase diagram discussed in Sec. V. Here PC indicates the 
compressible paramagnetic phase, FC the partially polarized 
compressible ferromagnetic phase, FSG the spin
gapped ferromagnetic phase, and NCF the non-collinear ferromagnetic phase. We emphasize that
this phase diagram is qualitative in nature and transition points vs. $\gamma$ should be taken as
upper limits.}
\label{phasediag} 
\end{figure}

We have used the Hartree-Fock approximation to study the 
competition between interactions and disorder near Landau
level filling factor $\nu=1$.  At a qualitative level our
results can be summarized by the schematic zero-temperature
phase diagram shown in Fig. \ref{phasediag}, which is drawn for the case of 
small but non-zero Zeeman coupling.  Distinct ground states can be 
distinguished by different values for the quantized Hall conductivity,
$\sigma_{xy}$, by the presence or absence of spontaneous spin-polarization
perpendicular to the direction of the Zeeman field, and by the presence
or absence of a gap for spin-flip excitations.  At small $\gamma$
(strong disorder), the electronic state is paramagnetic (denoted as PC in
Fig. \ref{phasediag}),
there is no spin-polarization in the absence of Zeeman coupling, and the 
Hall conductivity is expected to jump from $0$ to $2 e^2/h$ as the 
filling factor $\nu$ crosses the $\nu=1$ line.
For a small Zeeman coupling, there will
be a small splitting between the majority-spin and 
minority-spin extended state energies and the zero-temperature 
Hall conductance should have a narrow intermediate $e^2/h$ plateau centered on
$\nu=1$.  However, we do not expect that this plateau will be observable 
at accessible temperatures, and have indicated
this in Fig. \ref{phasediag} by using a thick line to mark the 
$0$ to $2 e^2/h$ phase boundary.  At somewhat larger $\gamma$ 
there is a phase transition at zero Zeeman energy between paramagnetic and 
ferromagnetic states (denoted as FC in
Fig. \ref{phasediag}).   In our calculations this transition occurs at a larger
value of $\gamma$ at $\nu=1$ than away from $\nu=1$.   As $\gamma$ increases
in the ferromagnetic state, we expect that the separation between
majority-spin and minority-spin extended state levels will increase
rapidly so that the $\nu=1$ integer quantum Hall plateau will
broaden and become observable.  At still larger $\gamma$, we find 
a transition to a state with the maximum spin polarization
allowed by the Pauli exclusion principle.  At $\nu \le 1$, this is full
spin-polarization.  In these states, marked `SG' for spin-gap in 
Fig. \ref{phasediag}, the differential spin-susceptibility
vanishes.  For realistic disorder models, it seems likely that in
the thermodynamic limit there will always be rare high-disorder
regions in the sample which prevent maximal spin-polarization from
being achieved.  For this reason, the phase transition we find 
in our finite systems likely indicates a crossover from  
large to small differential spin-susceptibility in macroscopic 
systems; we have therefore marked this transition by a dashed line.
Finally at the largest values of $\gamma$ (weakest disorder) 
the physics for $\nu$ near $1$ is dominated by Skyrmion quasiparticles
which emerge from the $\nu=1$ ferromagnetic vacuum.  In this regime,
the system develops spontaneous spin-polarization in the plane
perpendicular to the direction of the Zeeman field.  In Fig. \ref{phasediag}
we have labelled this regime NCF for non-collinear ferromagnet.

This phase diagram is intended to represent the filling factor interval
$0.85 \le \nu \le 1.15$, over which fractional quantum Hall effects 
are not normally observed and it appears likely that Hartree-Fock 
approximation calculations are able to represent interaction 
effects.  
Some of our findings may help explain the striking 
tilted plateau feature observed in the NMR spectra \cite{Barrett_private}
near $\nu=1$.
Nevertheless, we have found rich structure in the crossover between
non-interacting and disorder-free limits of the $\nu=1$ quantum
Hall effect which helps explain the difficulty experienced in
attempting to construct a simple interpretation 
of low-temperature NMR spectra. 
Our calculations motivate experimental studies of $\nu=1$ 
transport activation energy studies near the paramagnetic to 
ferromagnetic phase transition.

Helpful conversations with S.E. Barrett, Luis Brey, and  Tatsuya Nakajima 
are greatly acknowledged.  This work was supported by the National Science
Foundation under grants DMR-9714055 and DMR-9820816.

\end{document}